\documentclass[preprint,aps,pra, superscriptaddress, longbibliography]{revtex4-1}
\usepackage{amsmath}
\usepackage{amssymb}
\usepackage{graphicx}

\usepackage[
    colorlinks,
    linkcolor=blue,
    anchorcolor=black,
    citecolor=blue,
    urlcolor=blue,
]{hyperref}

\begin{document}


\title{Design and Fabrication of Metal-Shielded Fiber-Cavity Mirrors for Ion-Trap Systems}

\author{Wei-Bin Chen}
\affiliation{Laboratory of Quantum Information, University of Science and Technology of China, Hefei 230026, China}
\affiliation{Anhui Province Key Laboratory of Quantum Network, University of Science and Technology of China, Hefei 230026, China}
\affiliation{Hefei National Laboratory, University of Science and Technology of China, Hefei 230088, China.}

\author{Ding Fang}
\affiliation{Laboratory of Quantum Information, University of Science and Technology of China, Hefei 230026, China}
\affiliation{Anhui Province Key Laboratory of Quantum Network, University of Science and Technology of China, Hefei 230026, China}
\affiliation{Hefei National Laboratory, University of Science and Technology of China, Hefei 230088, China.}

\author{Cheng-Hao Zhang}
\affiliation{Laboratory of Quantum Information, University of Science and Technology of China, Hefei 230026, China}
\affiliation{Anhui Province Key Laboratory of Quantum Network, University of Science and Technology of China, Hefei 230026, China}
\affiliation{Hefei National Laboratory, University of Science and Technology of China, Hefei 230088, China.}

\author{Jin-Ming Cui}
\affiliation{Laboratory of Quantum Information, University of Science and Technology of China, Hefei 230026, China}
\affiliation{Anhui Province Key Laboratory of Quantum Network, University of Science and Technology of China, Hefei 230026, China}
\affiliation{CAS Center fro Excellence in Quantum Information and Quantum Physics, University of Science and Technology of China, Hefei 230026, China.}
\affiliation{Hefei National Laboratory, University of Science and Technology of China, Hefei 230088, China.}
\email{jmcui@ustc.edu.cn}

\author{Yun-Feng Huang}
\affiliation{Laboratory of Quantum Information, University of Science and Technology of China, Hefei 230026, China}
\affiliation{Anhui Province Key Laboratory of Quantum Network, University of Science and Technology of China, Hefei 230026, China}
\affiliation{CAS Center fro Excellence in Quantum Information and Quantum Physics, University of Science and Technology of China, Hefei 230026, China.}
\affiliation{Hefei National Laboratory, University of Science and Technology of China, Hefei 230088, China.}

\author{Chuan-Feng Li}
\affiliation{Laboratory of Quantum Information, University of Science and Technology of China, Hefei 230026, China}
\affiliation{Anhui Province Key Laboratory of Quantum Network, University of Science and Technology of China, Hefei 230026, China}
\affiliation{CAS Center fro Excellence in Quantum Information and Quantum Physics, University of Science and Technology of China, Hefei 230026, China.}
\affiliation{Hefei National Laboratory, University of Science and Technology of China, Hefei 230088, China.}
\email{cfli@ustc.edu.cn}

\author{Guang-Can Guo}
\affiliation{Laboratory of Quantum Information, University of Science and Technology of China, Hefei 230026, China}
\affiliation{Anhui Province Key Laboratory of Quantum Network, University of Science and Technology of China, Hefei 230026, China}
\affiliation{CAS Center fro Excellence in Quantum Information and Quantum Physics, University of Science and Technology of China, Hefei 230026, China.}
\affiliation{Hefei National Laboratory, University of Science and Technology of China, Hefei 230088, China.}

\date{\today}

\begin{abstract}
    Trapped ions in micro-cavities constitute a key platform for advancing quantum information processing and quantum networking. By providing an efficient light–matter interface within a compact architecture, they serve as highly efficient quantum nodes with strong potential for scalable quantum network. However, in such systems, ion trapping stability is often compromised by surface charging effects, and nearby dielectric materials are known to cause a dramatic increase in the ion heating rate by several orders of magnitude. These challenges significantly hinder the practical implementation of ion trap systems integrated with micro-cavities. To overcome these limitations, we present the design and fabrication of metal-shielded micro-cavity mirrors, enabling the stable realization of ion trap systems integrated with micro cavities. Using this method, we constructed a needle ion trap integrated with fiber Fabry–Pérot cavity and successfully achieved stable trapping of a single ion within the cavity. The measured ion heating rate was reduced by more than an order of magnitude compared with unshielded configurations. This work establishes a key technique toward fully integrated ion–photon interfaces for scalable quantum network.
\end{abstract}
\maketitle

\section{Introduction}
In quantum networks, efficient quantum interfaces require strong coupling between static and flying qubits \cite{kimble2008quantum}. Numerous experiments have identified photons as ideal flying qubits \cite{krutyanskiy2024multimode, walker2018long, walker2020improving, ward2022generation, krutyanskiy2023entanglement, schupp2021interface, brekenfeld2020quantum}, while placing static qubits inside optical cavities can greatly enhance their interaction with photons \cite{duan2010colloquium, reiserer2015cavity, reiserer2022colloquium}. Consequently, extensive studies have been conducted on the interaction between various materials and optical cavities \cite{stute2012tunable,keller2004continuous,cetina2013one, krutyanskiy2023entanglement, schupp2021interface, benedikter2017cavity, junge2013strong, jensen2014cavity, barclay2009coherent, krutyanskiy2023telecom, keller2004calcium, liu2012optogenetic, casabone2013heralded}.

The fiber Fabry–Pérot cavity (FFPC), serving as an open microcavity, provides a compact and highly integrable platform for light–matter interaction. Owing to its fiber-coupled interfaces, photons can be directly collected and transmitted through optical fibers. The reduced mode volume enhances the coupling strength between the ion and the cavity field, thereby facilitating high-bandwidth quantum repeater implementation. Advances in microfabrication techniques \cite{hunger2010fiber} have enabled successful coupling of diverse emitters to FFPC, including nitrogen–vacancy (NV) and silicon–vacancy (SiV) \cite{albrecht2014narrow, janitz2020cavity, haussler2019diamond}, quantum dots \cite{miguel2013cavity}, neutral atoms \cite{kato2015strong, gallego2018strong, uphoff2016integrated, wang2024ultrafast, wang2025purcell, niemietz2021nondestructive}, and trapped ions \cite{teller2023integrating, takahashi2020strong, ballance2017cavity}. Among these, trapped ions stand out as excellent stationary qubits with long coherence time and high-fidelity control \cite{wang2021single, sotirova2024high, erickson2022high}, while being stable trapping without reloading. Ion–FFPC integrated system (IFIS) thus combines the ions' superior coherence and controllability with efficient light collection and scalable interfacing, providing a promising route toward quantum network nodes with high efficiency.

However, two main challenges arise in the IFIS: (i) stray ultraviolet light causes charge accumulation on the fiber surface mirror \cite{harlander2010trapped}, leading to ion-trapping instability \cite{brandstatter2013integrated}; and (ii) the nearby dielectric mirrors can increase the ion heating rate by several orders of magnitude \cite{teller2021heating}. A common approach is to surround the fiber with a metallic shield, such as a metal tube, which has enabled stable ion–cavity coupling \cite{steiner2013single, takahashi2013integrated, takahashi2017cavity, takahashi2020strong, christoforou2020enhanced, fernandez2021fully, steiner2014photon, ballance2017cavity}. Nevertheless, this approach still faces several practical challenges, including limited fiber alignment precision \cite{kassa2018precise}.
Here, we present a fabrication technique for metal-shielded fiber-cavity mirrors that preserves optical transmission at the fiber end face while forming a metallic mask on both the fiber side and the outer annular region of the end face to maximize shielding. Compared with the metal-tube method \cite{kassa2025integrate}, our design exposes a smaller fiber area, providing stronger suppression of electric-field noise. Moreover, the gold layer exhibits excellent thickness uniformity and reproducibility, enabling fiber alignment with precision on the order of a few micrometers. We demonstrate that this structure effectively reduces charge accumulation and significantly suppresses dielectric-induced ion heating, as confirmed by finite-element simulations following \cite{teller2021heating}.
We further characterized the cavity finesse before and after processing, finding negligible degradation for cavity lengths up to $150\:\mu\text{m}$  and acceptable performance at $250\:\mu\text{m}$ . Using this fiber electrode in a needle-trap system, we have achieved stable single-ion trapping. The measured heating rate is consistent with the simulation results.

\section{Design and Simulation}
In the IFIS, the FFPCs primarily influence the trapped ions in two key ways: (i) the accumulation of positive charge on the fiber surface due to photoelectric effect, and (ii) the presence of nearby dielectric layers can significantly increase the ion heating rate. In the following sections, we model and analyze each of these effects separately, and discuss how our process mitigates both issues.\

\subsection{light-induced charge}
Since the cooling and ionization light for the majority of ions is ultraviolet or near-ultraviolet, irradiation on the optical fiber surface induces the photoelectric effect, causing electrons to be emitted from the valence band of the SiO2 material on the fiber surface, leading to the accumulation of positive charge. Due to the fiber's extremely low electrical conductivity ($10^{-14} \sim 10^{-18} \text{ S/m}$), the accumulated positive charges cannot be efficiently conducted away, leading to continuous charge buildup on the fiber surface. As the charge accumulates, the electrical potential on the fiber surface increases. Although the fiber has very low conductivity, it is not entirely non-conductive. As the electrical potential on the fiber surface rises, the current needed to conduct away the accumulated charge also increases. Once this current reaches a balance with the rate at which charge accumulates, a dynamic equilibrium is reached, and the charge accumulation on the fiber surface ceases. Assuming that scattered ultraviolet light is uniformly distributed in space, the irradiance on the fiber surface per unit area is constant, meaning the rate of positive charge generation per unit area is also constant. This rate can be expressed as a surface current density $i^+$. The total current is then given by $I^+=\iint_S i^+ \, \mathrm{d}\sigma$, where $S$ represents the total exposed surface area of the fiber. Based on measurements in \cite{ong2020probing}, it takes approximately one week for the fiber to accumulate charge up to $4\:\text{e}\!\cdot\!\mu\text{m}^{-2}$, leading to an estimated surface current density of $i^+ \sim 10^{-12} \mathrm{A\!\cdot\!m^2}$.

\begin{figure*}[!tbp]
    \centering\includegraphics[width=\textwidth]{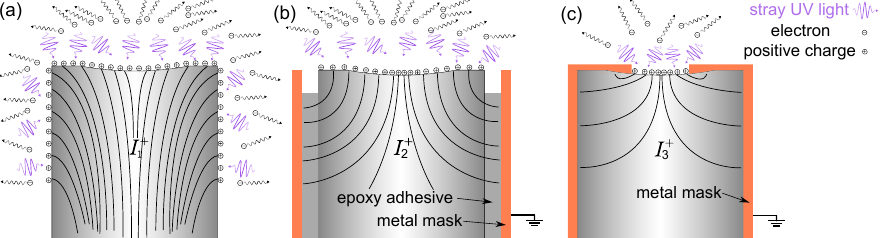}
    \caption{Photoelectric effect. (a) The optical fiber is fully exposed. (b) The optical fiber is enclosed in a metal tube, with gaps filled using thermosetting adhesive. (c) A gold layer is deposited on the fiber surface, exposing only the central $60\:\mu\text{m}$ region of the end face.}
    \label{photoelectric}
\end{figure*}

Three scenarios are considered: no shielding, the fiber is enclosed in a metal tube, and the fiber's end face center is exposed, as illustrated in Fig.~\ref{photoelectric}. In Fig.~\ref{photoelectric}(a), both the fiber's end face and side surfaces accumulate charge, and these charges can only be conducted away through the long fiber, resulting in a significant resistance. Additionally, the larger exposed surface area leads to a higher current $I^+$, which causes the electrical potential on the fiber surface to rise substantially. In Fig.~\ref{photoelectric}(b), the fiber is enclosed in a metal tube, with the fiber fixed within the tube using optical thermosetting adhesive. This configuration ensures that no charging occurs on the fiber's side and accelerates the dissipation of accumulated charges from the fiber surface, as they no longer need to traverse the long fiber. As a result, the electrical potential on the fiber surface decreases by several orders of magnitude compared to Fig.~\ref{photoelectric}(a). In Fig.~\ref{photoelectric}(c), a gold layer is deposited on the fiber surface, exposing only the essential central optical region of the end face, thereby maximizing the fiber's shielding. The exposed area is reduced further to one-quarter of that in Fig.~\ref{photoelectric}(b), and the path required for the conduction of these charges is thereby shortened. Therefore, Fig.~\ref{photoelectric}(c) results in a further reduction in the electrical potential on the fiber surface.

\begin{figure*}[!tbp]
    \centering\includegraphics[width=\textwidth]{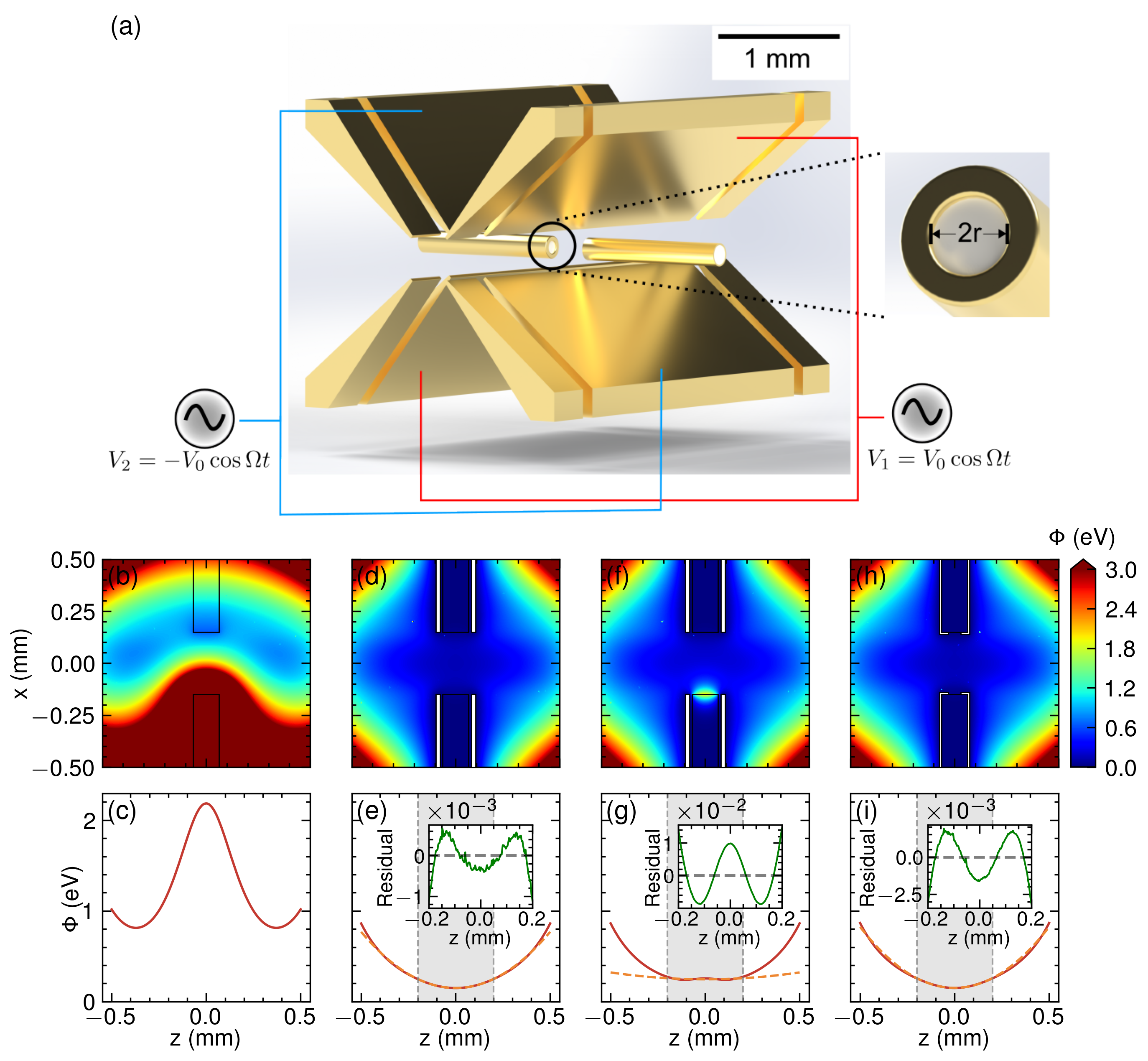}
    \caption{(a) Trap geometry used in the simulations. The optical fiber is aligned along the z-axis within a blade trap. The exposed end-face region has a diameter of $2r$, with $r = 30\:\mu\text{m}$ in this simulation. RF signals of opposite phase are applied to the two pairs of RF electrodes. Only the configuration corresponding to Fig.~\ref{photoelectric}(c) is shown; cases involving a metal tube enclosure or no surface metal shielding are not depicted. (b) Pseudopotential distributions for the unshielded fiber, the fiber enclosed in a metal tube (d, f), and the fiber with only the central $60\:\mu\text{m}$ of the end face exposed (h). The surface current density is $10^{-12}\,\mathrm{A{\cdot}m^{-2}}$ for (b, d), and $10^{-10}\,\mathrm{A{\cdot}m^{-2}}$ for (f, h). (c, e, g, i) Corresponding pseudopotential profiles along $x = 0$ for (b, d, f, h), respectively. Insets in (e, g, i) show quadratic fitting errors within the central $\pm\,200\:\mu\text{m}$ region.}
    \label{dianshimoni}
\end{figure*}

To investigate the influence of this surface current on the pseudopotential distribution, we modeled a blade trap and optical fiber cavity, simulating the pseudopotential distribution within the trap using finite element software, as illustrated in Fig.~\ref{dianshimoni}(a). A blade trap is selected because, in the case of a needle trap, it would be necessary to apply an Radio Frequency (RF) signal to the metal shielding layer outside the optical fiber. However, in subsequent simulations, the shape of the metal shielding layer at the end face of the optical fiber varies, which would lead to differences in the distribution of the pseudopotential. To investigate the effect of the charging current in the exposed region of the optical fiber on the pseudopotential, it is required to vary the exposed area of the optical fiber under the condition of a consistent pseudopotential distribution. Therefore, although we apply an RF signal to the fiber electrode in subsequent experiments, in the simulation conducted here, we choose to insert an optical fiber with a metal shield biased with a Direct Current (DC) voltage into the blade trap.

The trapping potential $\Phi_{\mathrm{RF}}$ generated by the RF electrodes is the time-averaged equivalent pseudopotential produced by the alternating voltage $U(\vec{r},t)=U_{\mathrm{RF}}(\vec{r})\cos(\Omega_{\mathrm{RF}} t)$ applied to the RF electrodes. Its expression is given by

\begin{equation}
    \Phi_{\mathrm{RF}}(\vec{r})=\frac{e^2}{4m\Omega_{\mathrm{RF}}^2}|\nabla U_{\mathrm{RF}}(\vec{r})|^2,
    \label{yanshi}
\end{equation}
where $e$ is the unit charge, $m$ is the mass of $^{138}\mathrm{Ba}^+$ ion, and $\Omega_{\mathrm{RF}}$ is the angular frequency of the RF signals. To obtain the pseudopotential distribution generated by the RF electrodes, we apply a voltage $U_{\mathrm{RF}}$ to the electrodes in the finite element software and set the boundary conditions, where other electrodes and infinite space are at zero potential. The resulting potential distribution $U_{\mathrm{RF}}(\vec{r})$ can then be used to calculate the corresponding equivalent pseudopotential distribution $\Phi_{\mathrm{RF}}(\vec{r})$ using the Eq. \ref{yanshi}. For the DC electrodes, the potential distribution must be computed separately. For the $i$-th DC electrode, we set the potential of all other electrodes and infinite space to zero, resulting in the potential distribution $\Phi_{\mathrm{DC},i}(\vec{r})$ produced by the $i$-th electrode in space. The potential due to the surface current on the fiber is calculated similarly to the DC electrode potential. Once the surface current is set, we set the potentials of all other electrodes and infinite space to zero and solve for the potential distribution $\Phi_{\mathrm{current}}(\vec{r})$ due to the surface current on the fiber. The total confinement potential experienced by the ion is then given by

\begin{equation}
    \Phi(\vec{r})=\Phi_{\mathrm{RF}}(\vec{r})+\sum\limits_{i}\Phi_{\mathrm{DC},i}(\vec{r})+\Phi_{\mathrm{current}}(\vec{r}).
\end{equation}

The fiber is modeled as a cylinder with a diameter of$125\:\mu\text{m}$ and a length of $60\:\text{mm}$, with the metal tube in Fig.~\ref{photoelectric}(b) having an inner diameter of $150\:\mu\text{m}$ and an outer diameter of $190\:\mu\text{m}$. In Fig.~\ref{photoelectric}(c), the thickness of the metal shielding layer is $10\:\mu\text{m}$, and the central exposed area has a diameter of $60\:\mu\text{m}$. The four central blades, each $2.2\:\text{mm}$ wide, are loaded with an RF signal, with opposite pairs of blades applying the signal with a $180^\circ$ phase difference. In the simulation, $V_0 = 30\:\text{V}$, $\Omega_{\text{RF}} = 2\pi \times 20\:\text{MHz}$, and the voltage applied to the eight Endcap electrodes is $V_{\text{endcap}} = 100\:\text{V}$. The resulting pseudopotential distribution is shown in Fig.~\ref{dianshimoni}. In Fig.~\ref{dianshimoni}(a), a surface current density of $10^{-12}\,\text{A}\!\cdot\!\text{m}^{-2}$ is applied to the fiber's end face and the side surface within $200\:\mu\text{m}$ from the fiber's tip. In Fig.~\ref{dianshimoni}(c), the surface current density is applied only to the central exposed region of the fiber's end face. Fig.~\ref{dianshimoni}(b, d) show the pseudopotential distribution through $x = 0$ for Fig.~\ref{dianshimoni}(a, c), respectively. As seen in Fig.~\ref{dianshimoni}(b), without the metal shielding, the surface current on the fiber distorts the pseudopotential into a double-well shape, with the lowest potential point deviating from the center by nearly $400\:\mu\text{m}$, and the potential barrier in the center reaching 1.4\:\text{eV}. When the fiber is enclosed in a metal tube, as shown in Fig.~\ref{dianshimoni}(d), the pseudopotential distribution can be well approximated by a quadratic function, with the fitting error within 1.5\:\text{meV} for the $\pm\,200\:\mu\text{m}$ region.

In Fig.~\ref{dianshimoni}(e), the fiber is enclosed in a metal tube, and the surface current density is increased by two orders of magnitude to $10^{-10} \, \mathrm{A\!\cdot\!m^{-2}}$. As seen in Fig.~\ref{dianshimoni}(f), the pseudopotential distribution forms a double-well shape. In Fig.~\ref{dianshimoni}(g), the fiber is maximally shielded, exposing only the central $60\:\mu\text{m}$ area of the end face with a surface current density of $10^{-10} \, \mathrm{A\!\cdot\!m^{-2}}$. From Fig.~\ref{dianshimoni}(h), it can be seen that the fitting error in the $\pm\,200\:\mu\text{m}$ range is only 4\:\text{meV}, effectively shielding the effect of the surface current density of $10^{-10} \, \mathrm{A\!\cdot\!m^{-2}}$.

\subsection{heating rate}

In ion trap systems, the primary source of ion heating is electric field noise \cite{turchette2000heating, brownnutt2015ion}. Ions undergo secular motion within the potential well\cite{leibfried2003quantum}, causing changes in the electric field distribution created by the ions. These variations in the electric field lead to modifications in the surface charge distribution on both the electrodes and the dielectric. When a dielectric is positioned near the ion, or when the ion-dielectric distance is comparable to the ion-electrode distance, the energy required to alter the charge distribution on the dielectric surface is significantly higher than that needed to modify the charge distribution on the electrode. According to the fluctuation-dissipation theorem (FDT) \cite{callen1951irreversibility}, the energy dissipated by the ion into the surrounding environment is equal to the energy exerted by the environment's reaction on the ion. This implies that when a dielectric is present, the heating power experienced by the ion is significantly greater than in the absence of the dielectric. This is the primary source of ion heating in the IFIS.

\begin{figure}[htbp]
    \centering\includegraphics[width=0.5\textwidth]{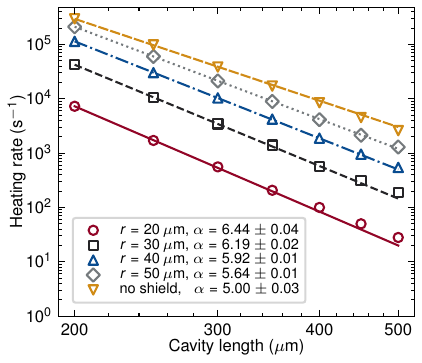}
    \caption{Simulation results for the relationship between the heating rate and the cavity length. The uppermost line corresponds to the case without a metal shield, and r decreases progressively from top to bottom.}
    \label{heating_rate_simulation}
\end{figure}
We use the method proposed in \cite{teller2021heating} to simulate the heating rate of our system. $d$ is the distance between the ion and the fiber end face, which also corresponds to the cavity length of $2d$. The electric field noise power spectrum, $S(\omega)$, which is due to the dissipation in the dielectric at temperature T, can be written as \cite{brownnutt2015ion}

\begin{equation}
    S(\omega_i) = \frac{4k_BT}{\Delta^2e^2\omega_i}\epsilon_0\epsilon_r\tan\delta\int_V|\vec{\Delta E_i}|^2,
\end{equation}
where $\vec{\Delta E_i}$ denotes the vector difference in the electric field between the ion at the origin and at a displacement of $\delta_i=\Delta$ in the $i$ direction, where $i=x, y, z$. $\omega_i$ is the ion's secular motion angular frequency, which is also the phonon frequency in different directions. $k_B$ is the Boltzmann constant, $\epsilon_0$ is the vacuum permittivity, $V$ is the dielectric volume, $\epsilon_r$ is the relative permittivity, and $\tan \delta$ is the loss tangent of the material. The ion heating rate is given by \cite{brownnutt2015ion}

\begin{equation}\label{eq:heatingrate}
    \dot{n_i} = \frac{e^2}{4m\hbar\omega_i}S(\omega_i).
\end{equation}

In the simulations presented in this paper, the phonon frequency along the fiber axis is $2\pi \times 2.85\:\text{MHz}$ \cite{fang2025cavity}, and the phonon frequency perpendicular to the fiber axis is $2\pi \times 1.3\:\text{MHz}$. Throughout this work, the heating rate discussed refer to the ion’s heating rate along the cavity axis. For the IFIS, the cavity mode forms a standing wave along the cavity axis with a period corresponding to half the optical wavelength, typically on the order of hundreds of nanometers—much smaller than the mode diameter, which is on the micrometer scale. The heating rate directly determines the ion’s temperature and motional amplitude, which in turn influence the ion–cavity coupling strength. Consequently, the ion's heating rate along the cavity axis has a significant impact on the coupling strength than those along the two orthogonal directions. Therefore, in this work we focus exclusively on the heating rate along the cavity axis.

It has been shown that when the ion is located at a distance $d$ from an infinite dielectric plane, $\dot{n} \propto d^{-\alpha}$, with $\alpha = 3$ \cite{brownnutt2015ion}. However, since the fiber end face is not an infinite dielectric plane, the exponent $\alpha$ becomes $4.016(6)$ \cite{teller2021heating}.When the fiber is enclosed in a metal tube, the exponent becomes $\alpha = 4.4$ \cite{kassa2025integrate}. A larger value of $\alpha$ indicates that the heating rate decreases more rapidly as the ion-dielectric distance increases.

We simulated the dependence of the heating rate on the cavity length for various optical region diameters $2r$, as illustrated in Fig.~\ref{heating_rate_simulation}. A smaller $r$ corresponds to a reduced exposed area of the fiber cavity mirror, which leads to a substantial increase in the power-law exponent $\alpha$. Our fabrication process yields an exponent of $\alpha = 5.98$ when $r = 30\:\mu\text{m}$, compared to $\alpha = 4.4$ obtained in the metal-tube enclosure approach \cite{kassa2025integrate}. This steeper scaling behavior indicates that the heating rate in our method decreases more rapidly with increasing cavity length. Consequently, our process provides a significantly stronger suppression of dielectric-induced heating effects, particularly as the cavity length $L$ increases.

\section{Method and Experiment}

\begin{figure*}[htbp]
    \centering\includegraphics[width=\textwidth]{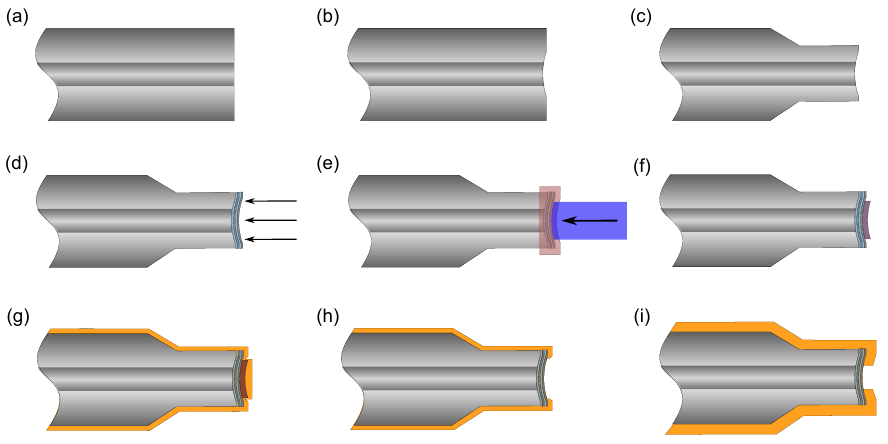}
    \caption{Flow chart for fabricating fiber electrodes. (a) Removing the coating of the fiber. (b) Creating concave surface by arc discharge. (c) Etching the fiber head with HF solution. (d) IBS coating. (e) Photolithography. (f) Developing. (g) Sputtering gold. (h) Photoresist strpping. (i) Electroplating thickening.}
    \label{process}
\end{figure*}
The experiment utilized a LMA-8 photonic crystal fiber (PCF), characterized by a central microstructure of air holes running along its length. Given that subsequent processing involves various chemical liquids, the ingress of these liquids into the internal air channels via the end-face poses a critical risk. Even if thermal treatment could be employed to clear the bulk liquid from the fiber channels, the complete removal of crystallization residues or other impurities remaining after chemical evaporation cannot be guaranteed. Such residual contaminants would detrimentally impact light transmission through the fiber. To mitigate this, we employed an arc discharge method. This technique serves a dual purpose: (i) it effectively seals and protects the internal hole region of the fiber, and (ii) it enables the formation of a concave surface suitable for the fiber cavity (as illustrated in Fig.~\ref{process}(b) and referenced in \cite{fang2024arc}).

The LMA-8 fiber possesses an initial head diameter of $240\:\mu\text{m}$. If the designed cavity length is maintained at approximately $250\:\mu\text{m}$, the resulting geometry restricts the optical access at the fiber side surface, limiting the light transmission angle to within approximately $45^\circ$. This severe angular constraint significantly complicates the arrangement of cooling light delivery and other necessary optical paths. To enhance this optical access, the fiber head must undergo hydrofluoric acid (HF) etching following the arc discharge treatment, thereby reducing the diameter from $240\:\mu\text{m}$ to approximately $120\:\mu\text{m}$, as illustrated in Fig.~\ref{process}(c).

Following the etching process, a high-reflective (HR) coating is deposited onto the fiber end face. Subsequently, the fiber end face is subjected to photolithography to pattern a photoresist layer. This process leaves a photoresist mask with a $100\:\mu\text{m}$ diameter. The photoresist employed for this application is AZ5214E photoresist, which was utilized in an image-reversal process to achieve negative-tone patterning. In the following part, we will describe each fabrication step in detail.

First, a pre-bake step is performed by placing the sample fiber on a $80\:^\circ\text{C}$ hotplate for 1 minute. This step helps evaporate the solvent from the photoresist, improving its adhesion to the fiber end face. Next, an exposure step is conducted, as illustrated in Fig.~\ref{process}(e). A $405\:\mu\text{m}$ laser is coupled into a PM-S405-XP fiber, which serves as the exposure light source for photolithography, referred to as the exposure fiber. The exposure fiber and the sample fiber are placed in a fiber splicer to achieve precise alignment and control of the end-face distance between the two fibers. The exposure fiber delivers $9\:\mu\text{W}$ of $405\:\mu\text{m}$ light power, producing a $15\:\mu\text{m}$-diameter spot at the sample fiber end face, which is positioned approximately $100\:\mu\text{m}$ away from the exposure fiber and the exposure time is 1.5\:s.

\begin{figure*}[!tbp]
    \centering\includegraphics[width=0.67\textwidth]{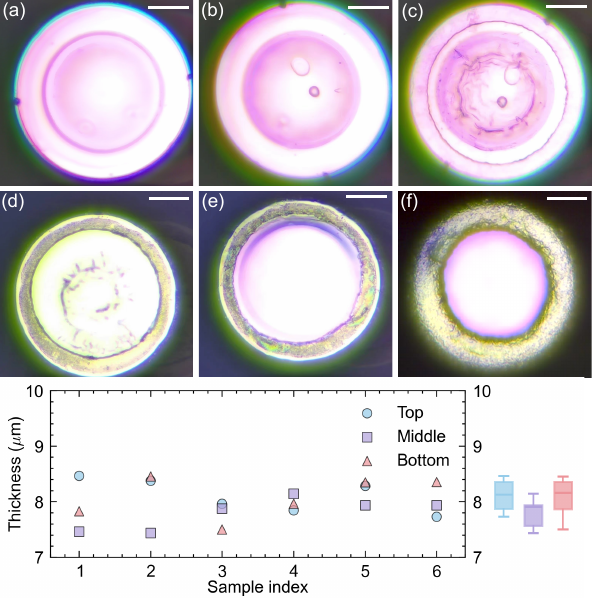}
    \caption{Schematic of the fabrication process (scale bar: $30\:\mu\text{m}$) and the corresponding thickness statistics. (a, b) End facet before photolithography and after reversal baking, respectively. (c)End facet after development, showing a circular photoresist region at the center serving as a sputtering mask. (d) After magnetron sputtering, a metal layer covers the entire end facet, with the central circle corresponding to the photoresist. (e) End facet after photoresist removal, revealing the central cavity mirror while retaining a surrounding metal ring. (f) After electroplating, the outer metal ring thickens and slightly extends toward the center. (g) Under identical electroplating parameters, multiple samples were fabricated, and the gold thickness was measured at various positions.}
    \label{process_real}
\end{figure*}

Following the exposure, a reversal bake is performed by placing the sample fiber on a $120\:^\circ\text{C}$ hotplate for 1 minute. This process renders the photoresist in the initially exposed region insoluble during development, while the unexposed outer ring remains photosensitive. Subsequently, a flood exposure is carried out—this step resembles a standard photolithography procedure and does not impose strict requirements on exposure intensity or duration, provided that the total exposure dose is sufficient. The purpose of this step is to expose the previously unexposed outer ring, enabling its removal during the subsequent development process. Development is then performed for 2 minutes to dissolve and remove the outer annular regions that have only undergone flood exposure, as illustrated in Fig.~\ref{process}(f) and \ref{process_real}(c). Finally, the fiber tip is briefly immersed in deionized water to eliminate any residual developer on the fiber surface, thereby preventing contamination that could affect the following sputtering step.

Through these steps, we can leave a photoresist mask in the $100\:\mu\text{m}$ central region of the fiber end face, which will serve as the mask for subsequent processes. Following the photolithography process, a 20\:\text{nm} titanium adhesive layer is first deposited using magnetron sputtering, followed by a 100\:\text{nm} gold layer, as illustrated in Fig.~\ref{process}(g) and \ref{process_real}(d).

Following magnetron sputtering, a thin gold film is deposited across the exposed fiber surface. However, due to the inherently poor adhesion of sputtered metals to the photoresist, the gold layer on the central $100\:\mu\text{m}$ region is only weakly bonded. During the lift-off process, the fiber sample is immersed in an acetone bath and subjected to ultrasonic agitation. The induced cavitation and fluid motion facilitate the removal of the weakly adhered gold film, thereby exposing the underlying photoresist layer. The exposed photoresist is then rapidly dissolved by acetone, leaving the $100\:\mu\text{m}$ central optical region of the fiber end face uncovered. In contrast, the gold layer deposited directly onto the outer ring region—where no photoresist mask was present—remains strongly adhered to the fiber surface and is unaffected by the ultrasonic agitation, as illustrated in Fig.~\ref{process}(h) and \ref{process_real}(e).

The gold film deposited by magnetron sputtering has a thickness on the order of $100\:\mu\text{m}$, which is insufficient for use as a dedicated radio-frequency (RF) electrode in the ion trap, as it is far below the material’s skin depth. At the operating RF frequency of approximately 20\:\text{MHz}, the skin depth of gold is about $15\:\mu\text{m}$. When the coating thickness is much smaller than this value, the effective resistance to the RF current increases substantially, resulting in pronounced Joule heating within the electrode and, consequently, an elevated ion heating rate in the trap. To mitigate this effect, the sputtered gold layer is subsequently thickened by electroplating until its thickness approaches the skin depth, as illustrated in Fig.~\ref{process}(i) and \ref{process_real}(f).

Under identical electroplating conditions, multiple samples were prepared. For each sample, the coating thickness was measured at the upper, middle, and lower sections within a 3\:\text{cm}-long plated region. The metal thickness was determined by subtracting the original fiber diameter of $240\:\mu\text{m}$ and dividing the difference by two, as summarized in Fig.~\ref{process_real}(g). The results show an average coating thickness of approximately $8\:\mu\text{m}$ with variations below $1\:\mu\text{m}$, demonstrating excellent thickness uniformity along individual fibers and high reproducibility across multiple electroplating runs.

\section{Result}

\subsection{Fiber Cavity Alignment}
\begin{figure}[!tbp]
    \centering\includegraphics[width=1\textwidth]{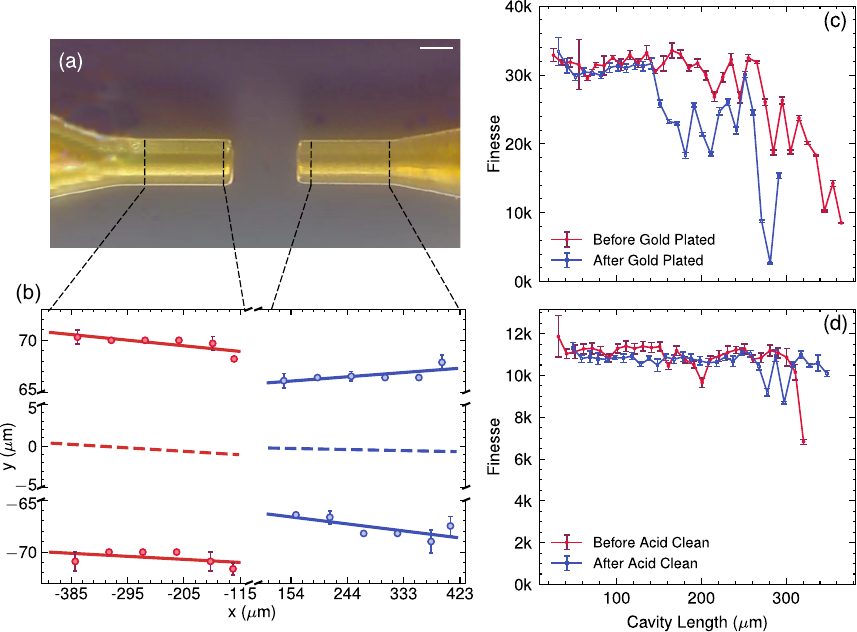}
    \caption{(a) Photograph of the fiber cavity with a scale bar corresponding to $100\:\mu\text{m}$. (b) Based on (a), the upper and lower edges of the left (red) and right (blue) fibers were identified (scattered points). The fiber edges were then obtained by fitting these points (solid lines), from which the fiber central axes were determined (dashed lines). The resulting alignment precision between the two fibers is $1.79 \pm 0.32\:\mu\text{m}$, with an angular deviation of $1.98 \pm 1.14\:\text{mrad}$. (c, d) Comparison of fiber cavity mirror finesse before and after (c) gold plated and (d) acid procedure.}
    \label{finesse_and_heatingrate}
\end{figure}

To evaluate the precision of the fiber-to-fiber alignment, the relative positions and orientations of the two fibers were characterized. Images were acquired along two orthogonal directions perpendicular to the fiber axis (only one direction is shown in Fig.~\ref{finesse_and_heatingrate}(a)). The edges of the fiber tips were identified, and four edge lines were fitted to determine the central axis of each fiber. From the two fitted axes, the positional and angular deviations between the two fibers were extracted, as illustrated in Fig.~\ref{finesse_and_heatingrate}(b). The analysis shows that the achieved alignment precision is on the order of $1\text{--}2\:\mu\text{m}$, with a negligible angular deviation.

\subsection{Finesse}

In our fabrication process, the fiber cavity mirrors are exposed to various chemical solution, including photoresist, developer solution, acetone, and gold electroplating solution. Although the outermost layer of the fiber cavity mirror is composed of SiO2, which should theoretically remain chemically inert to these substances, it cannot be guaranteed that the mirror finesse remains unaffected after undergoing these treatments. To evaluate this, a finesse measurement was performed on the fiber sample following gold electroplating, as illustrated in Fig.~\ref{finesse_and_heatingrate}(c). The measurement setup consisted of a flat mirror with identical coating parameters to the fiber cavity mirror and a one-dimensional piezoelectric translation stage (PI P-628) for scanning the cavity length \cite{fang2024arc}. As illustrated in Fig.~\ref{finesse_and_heatingrate}(c), the finesse remains nearly constant for short cavity lengths. However, when the cavity length increases to approximately $150\:\mu\text{m}$, a noticeable degradation in finesse is observed. We attribute this behavior to the expansion of the optical mode area on the cavity surface at larger separations, which enhances wavefront distortion losses and consequently reduces the cavity finesse.

Moreover, in cases where the electroplated fiber end face does not meet the experimental requirements and re-electroplating is necessary, the previously deposited metal layers must be completely removed. The gold layer is typically dissolved using aqua regia, while the underlying titanium adhesion layer is stripped with hot, concentrated hydrochloric acid. To evaluate the influence of this acid-cleaning process on the fiber cavity performance, we measured the finesse before and after the procedure, as illustrated in Fig.~\ref{finesse_and_heatingrate}(d). A decrease in finesse was observed following the acid-cleaning procedure. Specifically, for cavity lengths below $240\:\mu\text{m}$, the average finesse dropped from approximately 10800 to 10400, corresponding to an increase in optical loss of about 20\:\text{ppm}. We attribute this degradation primarily to residual metallic particles on the fiber end face that are difficult to completely remove.
\begin{figure}[!tbp]
    \centering\includegraphics[width=1\textwidth]{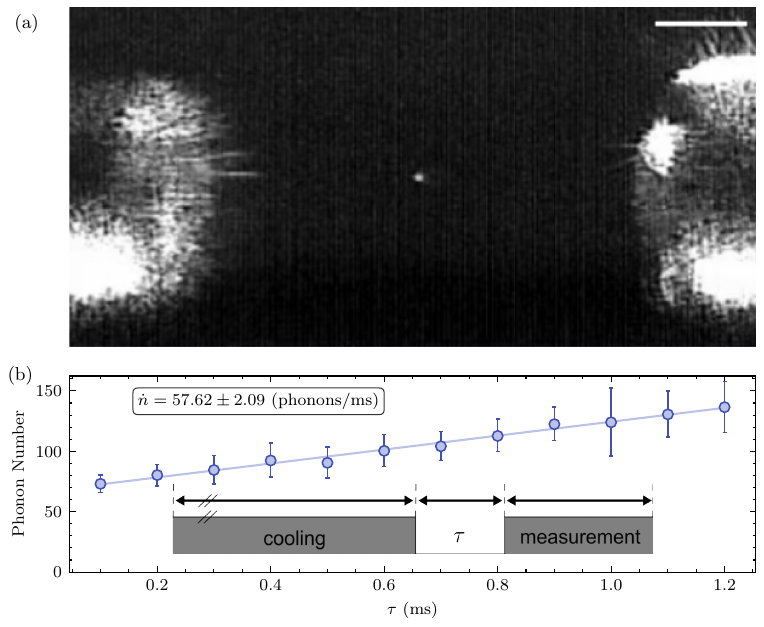}
    \caption{(a) Single ion trapped inside the fiber cavity with a scale bar corresponding to $50\:\mu\text{m}$. (b) Heating-rate measurement. The measured heating rate is 57.62\:\text{phonons/ms}. The inset shows the measurement sequence, where the delay time $\tau$ is varied to obtain the phonon number as a function of delay time, from which the heating rate is extracted by linear fitting. The ion temperature is determined from the contrast in the photon scattering rate into the cavity mode when the ion is positioned at different locations within the cavity standing-wave field \cite{fang2025cavity}.}
    \label{ion_in_cavity_heating_rate}
\end{figure}

\subsection{Heating Rate}

Using the fiber electrodes fabricated through the aforementioned process, we constructed a needle trap. Stable single-ion trapping was successfully achieved within a fiber cavity with a cavity length of $230\:\mu\text{m}$ as illustrated in Fig.~\ref{ion_in_cavity_heating_rate}(a), and a temperature measurement technique based on this system was developed \cite{fang2025cavity}. By measuring the ion temperature after various delay times of the cooling light, we obtained the relationship between the phonon number and the cooling-light pause duration, enabling us to quantify the heating rate of the system. The measured heating rate was $57.62 \pm 2.09\:\text{phonons/ms}$, as illustrated in Fig.~\ref{ion_in_cavity_heating_rate}(b). Using our experimental parameters in the previously described heating-rate simulation yielded a simulated value of 20.65\:\text{phonons/ms}, which is of the same order of magnitude as the experimental result.

In the case of a bare fiber cavity with a cavity length of $500\:\mu\text{m}$, the simulated heating rate is 95.04\:\text{phonons/ms} \cite{teller2021heating}. Using the relation $\dot{n} \propto d^{-\alpha}$ with $\alpha = 4.02$, the corresponding heating rate for a cavity length of $230\:\mu\text{m}$ is calculated to be 1810\:\text{phonons/ms}. When the fiber is enclosed in a metal tube \cite{kassa2025integrate}, the simulated heating rate at the same cavity length decreases to 72.1\:\text{phonons/ms}. It is generally observed that experimental heating rates exceed simulated values \cite{teller2021heating}. Nonetheless, our lithography and electroplating process reduces the heating rate by more than one order of magnitude compared to the bare fiber case and achieves superior performance relative to the metal-tube enclosure method.

\section{Conclusion}

We have demonstrated a fiber-electrode fabrication technique that enables a novel approach for constructing IFISs. Simulation results demonstrate that this method provides over two orders of magnitude improvement in suppressing surface currents on the fiber compared to the conventional approach of enclosing the fiber in a metal tube. Using this technique, we successfully built an ion trap system integrated with an FFPC and achieved long-term stable trapping of a single ion. The ion heating rate was subsequently measured, demonstrating effective shielding of the ion from dielectric-induced heating. Compared with the unshielded case, our approach reduces the ion heating rate by more than two orders of magnitude. Furthermore, it outperforms the metal-tube enclosure method. The finesse of the fiber cavity was characterized before and after the fabrication and acid-cleaning processes, confirming that the overall impact of these procedures remains acceptable within a practical range of cavity lengths. 

\section{Acknowledgement}

This work was supported by the National Key Research and Development Program of China (2024YFA1409403), the National Natural Science Foundation of China (Grants No. 11734015 and No. 12204455), the Innovation Program for Quantum Science and Technology (Grant No. 2021ZD0301604 and No. 2021ZD0301200), the Key Research Program of Frontier Sciences, CAS (Grant No. QYZDY-SSWSLH003), and the Anhui Provincial Natural Science Foundation, China (Grant No. 2508085MA005).
We also acknowledge the USTC Center for Micro and Nanoscale Research and Fabrication for partial support, and thank Wei Yu, He Yi-zhao, and Ma Jun for their technical assistance.


\providecommand{\newblock}{}

\end{document}